\documentclass[fleqn,twoside]{article}
\usepackage{espcrc2,epsfig}

\usepackage{graphicx}

\newcommand{\AmS}{{\protect\the\textfont2
  A\kern-.1667em\lower.5ex\hbox{M}\kern-.125emS}}

\title{A simple Lattice Model for hysteresis loops with exchange bias.}

\author{Eduard  Vives \address{
Departament d'Estructura  i Constituents de la  Mat\`eria, Universitat de  Barcelona, \\ Facultat de
F\'{\i}sica, Diagonal 647, 08028 Barcelona},   Xavier  Illa \addressmark  and   Antoni  Planes  \addressmark }

\begin{document}

\begin{abstract}
A simple lattice model that allows hysteresis loops with exchange bias
to be reproduced  is presented.  The model is  based on the metastable
Random  Field   Ising  model,  driven  by  an   external  field,  with
synchronous  local relaxation  dynamics.   The key  ingredient of  the
model is that a certain fraction $f$ of the exchange constants between
neighbouring spins is enhanced to a very large value $J_E$.  The model
allows the dependence of several properties of the hysteresis loops to
be analyzed as a function  of different parameters and we have carried
out an analysis of the first-order reversal curves.

\vspace{0.5mm}

\vspace{-1mm}

\end{abstract}
\maketitle

\begin{figure}[thb]
\includegraphics[width=7cm,clip]{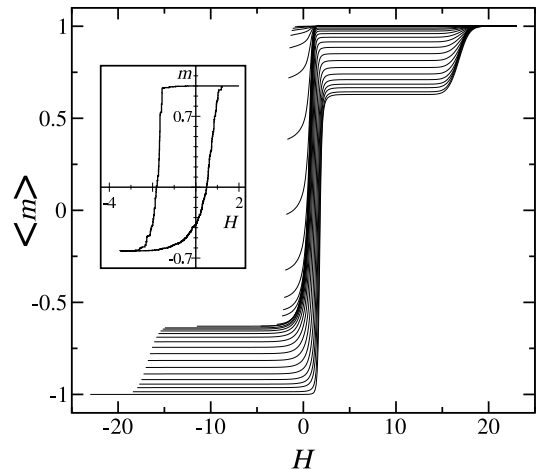}
\vspace{-6mm}
\caption{ A set of FORCS  for the EE-RFIM. They correspond to averages
over 1000  realizations with $\sigma$  = 1.0 and $f$=0.05.   The inset
shows an  example of  a hysteresis loop with  EB obtained with  the same
parameters, reversing the field at $H=-4$}
\label{fig1}
\vspace{-8mm}
\end{figure}

The  use of zero-temperature  lattice models  to study  hysteresis has
been  very  fruitful for  the  understanding  of  the role  played  by
disorder in determining loop  shape and Barkhaussen noise distribution
\cite{Sethna1993}.  Recently  the well known Random  Field Ising Model
(RFIM), has  been extended  in order to  account for the  existence of
exchange bias (EB). This property  consists in a shift along the field
axis of the hysteresis loop.  EB has been experimentally found in many
FM/AFM bilayer  systems \cite{Nogues1999}.  The key  ingredient of the
so  called Exchange  Enhanced  RFIM (EE-RFIM)  \cite{Illa2002} is  the
existence  of  a random  fraction  of  the nearest-neighbour  exchange
interactions (of the ferromagnetic layer)  which is enhanced to a very
large value.  The  Hamiltonian of the model on  a $N=L\times L$ square
lattice (that only represents the FM layer) can be written as:
\begin{equation}
{\cal H}  = -\sum_{ij}^{nn} J_{ij}  S_i S_j -  \sum_{i}^N h_i S_i  - H
\sum_{i}^N S_i
\label{hamiltonian}
\end{equation}
where $S_i=\pm  1$ are  the spin  variables on each  site, $H$  is the
external driving field, $h_i$ are quenched Gaussian random fields with
zero  mean  and  variance  $\sigma^2$.   The  exchange  constants  are
$J_{ij}=1$ (energy unit),  except for a random fraction  $f$ which has
an  enhanced  value  $J_{ij}=J_E$.   The  rate-independent  hysteresis
properties  of  this  model  can  be simulated  by  using  synchronous
me-tastable dynamics  \cite{Sethna1993}.  The inset  in Fig.~\ref{fig1}
shows

\noindent an example of  a hysteresis loop exhibiting  EB.  The
reason  behind  this EB  property  is that  such  loops  do, in  fact,
correspond  to partial  loops. The  symmetric loop  is  obtained after
applying very negative ($\propto J_E$) external fields.

In  this  short paper  we  present  an analysis of  the  First-Order
Reversal  Curves  (FORC) of  the  EE-RFIM  corresponding  to the  case
$J_E=20$  and  $f=0.05$  obtained  from  the simulation  of  a  $L=50$
lattice.   Other  interesting  properties  of  this  model  have  been
published elsewhere \cite{Illa2002}.

\begin{figure}[bht]
\includegraphics[width=7.5cm,clip]{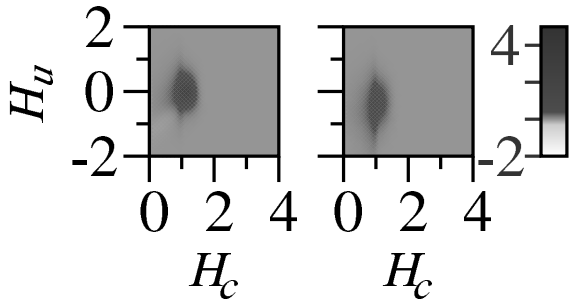}
\vspace{-9mm}
\caption{Averaged FORC diagram  of the RFIM (a) and the
EE-RFIM  (b) corresponding  to  $\sigma=2$.  The  scale  on the  right
indicates  the values of  $\rho$ corresponding  to different shades of grey.}
\label{fig2}
\end{figure}

\begin{figure}[t]
\includegraphics[width=7cm,clip]{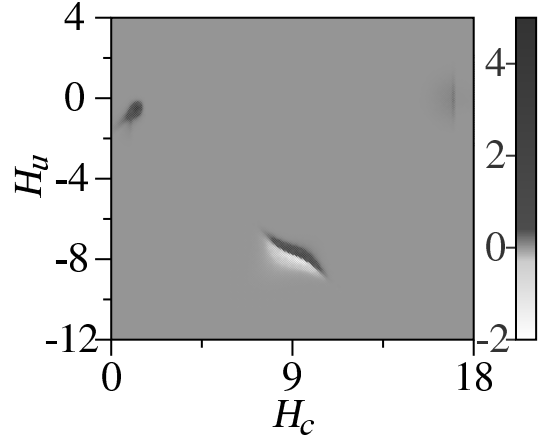}
\vspace{-10mm}
\caption{FORC diagram corresponding to the set of curves in Fig.~1.}
\label{fig3}
\vspace{-6mm}
\end{figure}

A set  of FORC is shown  in Fig.~\ref{fig1}.  In order  to analyze the
structure of  this set of FORC  curves, we compute the  so called FORC
diagram. This  is obtained by calculating the  mixed second derivative
of  the magentization  $m$  at field  $H_b$  on a  FORC obtained  with
reversal point $H_a$ \cite{Pike1999}:
\begin{equation}
\rho(H_a,H_b) = - \frac{\partial^2 m(H_a,H_b)}{\partial H_a \partial H_b} 
\end{equation}
The values  of $\rho$  are usually represented  on a diagram  which is
obtained  after rotating  the  co-ordinates according  to $H_c=(H_b  -
H_a)/2$ and $H_u = (H_a + H_b)/2$. The new co-ordinates allow a direct
comparison   with  the  Preisach   distributions  \cite{Bertotti1998}.
Fig.~\ref{fig2}   shows  the   central  part   of  the   FORC  diagram
corresponding to  the standard  RFIM (a) and  to the EE-RFIM  (b). The
comparison reveals that the EB property is reflected by a displacement
of the  central peak to the  bottom.  The vertical width  of this peak
increases  when the width  $\sigma$ of  the random  field distribution
also  becomes larger.  The  shift to  the bottom  is also  greater for
increasing $f$. Furthermore, other  interesting features appear on the
FORC  diagram.    These  can   be  observed  on   the  wider   map  in
Fig.~\ref{fig3}. The two main characteristics are the occurrence of an
oscillation in the  $H_u<0$ region and a second peak  in the region of
large values of $H_c$ ($\sim 17$).
\vspace{-6mm}
%
%
An interesting point to make is that the oscillation includes a region
of negative  values of $\rho$.   Neither this negative region  nor the
asymmetry of  Fig.~\ref{fig3} can be interpreted  within the framework
of the Preisach model for which  $\rho$ is assumed to be a probability
density  which  should always  be  positive  and  symmetric about  the
$H_u=0$ axis.  The negatives values  of $\rho$ are associated with the
fact that the  slope of the $m(H)$ curves at  $-1<H <0$ increases when
$H_a$ becomes larger.  It will  be very interesting to analyze whether
such singular  features occur for experimental systems  with EB.  This
will  reveal the  non-Preisach nature  of such  bilayered  systems. We
acknowledge fruitful comments from J.Nogu\'es and H.G.Katzgraber.

\end{document}